# Efficient Certificate Management in VANET

Ghassan Samara[#1], Wafaa A.H. Al-Salihy[*2], R. Sures[#3]

[#]National Advanced IPv6 Center, Universiti Sains Malaysia
Penang, Malaysia
[1]ghassan@nav6.org, [3]sures@nav6.org
[*]School of Computer Science, Universiti Sains Malaysia
Penang, Malaysia
[2]wafaa@cs.usm.my

*Abstract*—**Vehicular Ad hoc Networks is one of the most challenging research area in the field of Mobile Ad Hoc Networks, in this research We propose a flexible, simple, and scalable design for VANET certificates, and new methods for efficient certificate management , which will Reduce channel overhead by eliminating the use of CRL, and make Better certificate Revocation Management.**

**Also it will increase the security of the network and helps in identifying the adversary vehicle.**

## I. INTRODUCTION

VANET security has gained the most research efforts in the past years; Certificate plays an important role in network communication, any node in the network can't participate in the network without appropriate certificate.

Dealing with certificate management raises many other issues like certificate revocation list (CRL) that causes network overhead, a CRL is a list containing the serial numbers of all certificates issued by a given certification authority (CA) that have been revoked and have not yet expired. CRL makes overhead and expensive to use especially in high mobile network.

In this paper we concerned with certificate security and operations, how to protect system from adversary vehicles, how to distribute information about adversary vehicles, how to revoke certificates from adversary and assign new certificate for it, how to make secure connection, in sec. 2 we analyze the current research efforts in area of VANET certificates, in sec. 3 we are addressing our proposed network that contains solutions for current system.

## II. ANALYSIS OF RELEVANT RESEARCH AREA:

CRL is the most and common solution for certificate revocation and management, many papers tried to adapt the CRL solutions.

Efforts made by [1] aiming to reduce the CRL by using regional CA and using short lived certificates for traveling vehicles, the result obtained by the author " the distribution of CRL requires tens of minutes", which is too long time for a high and dense network like VANET, the authors in [2] proposed an idea to easy disseminate the CRL, by deploying C2C communication for distributing CRL, this will make faster distribution, but still CRL has a huge size and require time and processing complexity to search in, another work [3] , made many experiments on the size of CRL and how to distribute the CRL in the VANET network, the result says, when the size of CRL is high the delay time for receiving it will be high, another idea proposed in [4] says that CRL will store entries for less than a year old, this idea used to decrease the size of CRL, but still suffer from huge size, while Authors in [6] suggested a way to increase the search in CRL by using Bloom filter.

The authors in [10] proposed the use of Bloom filter to store the revoked certificate, and dedicate the CRL just to sign the revocation key for each vehicle, the use for Bloom filter will increase the speed for searching in it, but still the idea is to use the CRL, the problem of bloom filter as it is probabilistic function, and may give wrong information, as the certificate may not be in the list, and the result that the certificate is in the list.

The previous work and efforts didn't eliminate the Problems of CRL like Huge size, no central database for it, Channel overhead, communication overhead, processing overhead.

Authors in [5] introduce the use of temporary certificates and credentials for each geographic area so any new vehicle will not find any difficulty for communicating with current network, but this solution requires a dedicated work from CA to create and revoke certificate for each new coming and leaving vehicle.

Raya et al. [6] proposed a design of three novel protocols for revocation, RC2RL (Revocation using compressed Certificate Revocation Lists), RTPD (Revocation of the TPD), DRP (Distributed Revocation Protocol)  the first one tried to compress the CRL to solve the size of CRL, the second protocol tries to solve the revocation problem when the CA in presence, and the third when CA is not found, these protocols added more complexity to network and didn't deal with all cases of certificate management, while authors in [9] made analysis on the protocols of [6] and proposed new and more efficient revocation protocol called DRTA aims to use short lived certificates issued by CA.

Authors in [7] suggested a simple way for revocation, but it makes the channel vulnerable against attacks like Sybil attack, and causes a channel overhead.

The authors in [2] proposed the employment of vehicles to distribute CRL, when Road Side Unit (RSU) is not exist, this will reduce the communication bandwidth usage, but still has a problem as the relay from vehicle to vehicle raises security concerns.

Authors in [8] proposed two ways for revocation, but all of them are the responsibility of the vehicle itself, this giving the adversary an opportunity for avoiding the revocation, it is also mentioned that each vehicle must have a notion of trust.

## III. PROPOSED NETWORK

### A. The basic Idea:

- *Certificates:*

In normal network system, each vehicle must have a certificate for transmission, and this allows each vehicle to transmit even if it considered as adversary, thus it is a problem needs to be solved, many papers introduced ideas to solve it [1], [2], [3] and [6], and the most common idea is the use of Certificate Revocation List (CRL), CRL will keep the ability for the vehicle to transmit, if any vehicle receives information from a revoked vehicle it will accept the information and apply the id of the sender to the CRL, if the id in the list the receiver will ignore the message, otherwise it will take it, this procedure causes network overhead for frequent retransmission of CRL and causes high computation overhead for each vehicle when receiving any information, and again allows the adversary vehicle to transmit, in some situations the receiving vehicle may accept the information received from adversary, as not all vehicles have the updated CRL.

Our new idea is to provide each vehicle with special certificate; this certificate will insure the intention status of the vehicle, a Valid Certificate (VC) will be given to the valid vehicle (I mean: not adversary), and Adversary Certificate (AC) for adversary vehicle.

Each certificate will require 100 byte from memory, and the design of VC as in figure 1.

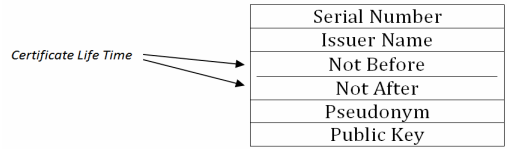

Fig. 1: Valid Certificate

And the design of AC as in figure 2.

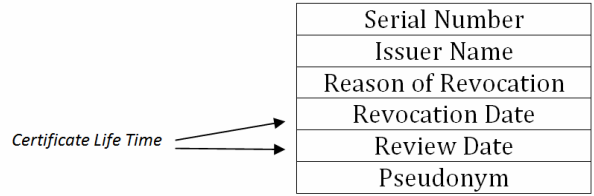

Fig. 2: Adversary Certificate

The use of this kind of certificates will eliminate the need for CRL.

The life time for AC will be one year, and to be checked by the traffic authority when making yearly checkup, as the problem maybe from the hardware.

The reason of revocation will follow this code:

| Code | Reason of Revocation |
|---|---|
| 1 | Bogus traffic information |
| 2 | Disruption of network operation |
| 3 | Cheating with identity, position or speed |
| 4 | Uncovering the identities of other vehicles |

Table 1: Table of Reason of Revocation.

Each vehicle has its own electronic document this document called certificate, this certificate has a limited and short life, authors in [10] suggested the use of each certificate for 10 minutes after that the certificate will expire, the use of these certificates is to insure the identity of the sender, VC certificate will insure the intention of the sender, both certificates will be attached along with every message, when a vehicle transmit a message, first it encrypts the message and the certificates and attach them together and send them, when a receiver receives a message, simply it will decrypt the certificate, if the certificate is VC, the receiver will decrypt the message and accept it, if the certificate is AC, the receiver will ignore the message, and makes a warning message for all the neighboring vehicles about that adversary vehicle, see figure 5.

The warning includes

| Warning Issuer Id |
|---|
| Adv. Id |
| Time Stamp |
| Reason of Revocation |
| AC Review Date |

Fig. 3: AC Warning

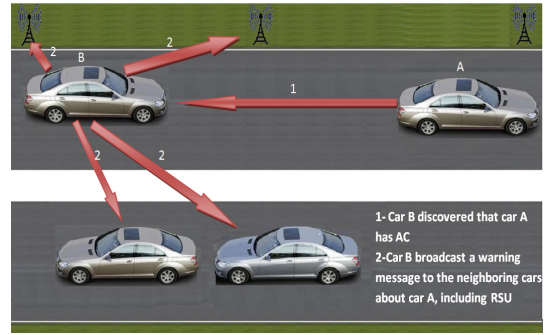

Fig. 5: AC Warning Distribution.

The "Adversary id" is the id of the adversary, "warning issuer id" used to know who made the warning, "time stamp" to insure the freshness of the warning, "reason of revocation" the code used to know why the certificate been revoked and "AC review Date" to know when this AC will expire and need to be reviewed.

The warning will be stored in a small list called Adversary List (AL), this list contains 10 entries, and each entry represents an adversary vehicle.

| Warning Issuer ID | Adv. Id | Time Stamp | Reason of Revocation | AC Review Date |
|---|---|---|---|---|
|  |  |  |  |  |
|  |  |  |  |  |

Fig. 4: Adversary List.

The importance of this list is that it is internal list, so no need to retransmit it periodically like CRL, the size of it is small, and of course smaller than the CRL is, it is fully updated as it contains information about current network and information about neighboring vehicles, as not all vehicles need revocation information about the whole network, if a vehicle wants to know information about the revocation information for the whole network, it simply makes request from nearest RSU available asking for CRL.

This list is ordered as new entry will be added at the top, and other nodes in the list will be stepped down, the tenth element will be removed first, the size of AL has a tradeoff, as if it is bigger, it will contain more adversary id's but the search in it will be slower, we have used 10 elements only as it contains the adversaries from current road, not all the world of VANET.

The conditions for removing an entry from the list are when new entry arrives, the tenth element will be removed, or when the vehicle leaves the street.

To take advantage from the list, we propose a new idea for receiving procedure, after the sender sends the encrypted message with the encrypted certificate, the receiver will take the id of the receiver and apply it to AL, if the id of the sender exists in AL, this means that the of the sender is not in the AL then the receiver will decrypt the certificate, if the certificate is AC the sender must be considered as new adversary, so the receiver will make sender is an adversary, so the receiver will ignore the message and it will move the id of the sender to the top of AL, if the id three steps, first it will add the senders id to AL, second step, make a warning message to the neighbors about this adversary, and third step, ignoring the message, if the certificate is VC then the receiver will take the message.

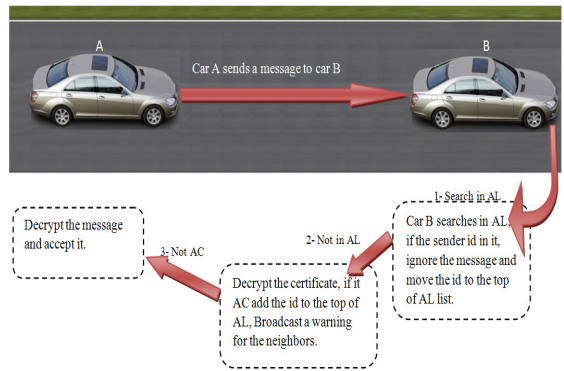

Fig. 6: Receiving New Message

We made the search in the list first as it is faster than the decryption, and this method is more efficient and faster than searching in CRL.

- *Revocation:*

Certificate revocation is used as a solution to degrade the adversary abilities from harming the system, the proposed protocols by [6] adds some complexity to the network, however, making an efficient and robust revocation is a hot research topic, in this paper we propose new protocol for revocation called Valid Adversary Protocol (VAP).

- *How it Works*

Before we start to talk about the protocol procedure we would like to introduce an idea for making category for every transmission, so any transmission of a message will be classified into a specific category, this will make it easy to the vehicle to track and analyze the received messages.

| Code | Priority | Application |
|------|----------|-------------|
| 001 | Safety of Life | Intersection Collision Warning /Avoidance |
| 002 | Safety of Life | Cooperative Collision Warning |
| 003 | Safety | Work Zone Warning |
| 004 | Safety | Transit Vehicle Signal Priority |
| 005 | Non-Safety | Toll Collection |
| 006 | Non-Safety | Service Announcement |
| 007 | Non-Safety | Movie Download( 2 hours of MPEG 1) |

Table 2: Message Category.

- *VBP Steps*

1) Step 1- suspicion: any vehicle suspect an action from a certain vehicle, the way that the vehicle sense the suspicion is out of the scope of the paper and will be considered as future work, but we can consider a case, as an example, where a vehicle receives information from 9 vehicles, all of them has VC and sending message with the same category and information, except of one of them sending a contradictory information, this vehicle could be a suspect, if this vehicle kept sending a contradictory information more than once, an accusation message will be sent to RSU.

2) Step 2- Accusation: Local RSU begin to receive many accusations about certain vehicle, the accusation (Ac) must be encrypted with the public key (PK) of RSU, RSU decrypt the accusation messages and make sure about VC of each sender, if number of accusers (AcV) is more than threshold, let us say the half number of current vehicles on the road, then RSU will consider the accused vehicle (AV) as an adversary, and it will send an accusation to CA.

   From AcV →

   Accusation = $Ac_{Kp}$ {RR, VC, Sig, TS, AV} (1)

   TS: Time stamp, Sig: Signature of the accuser, RR: Reason of Revocation, VC: Valid Certificate of the accuser, $_{Kp}$: public Key of RSU.

3) Step 3- RSU will send the accusation message to CA, and CA will consider the accused vehicle as an adversary.

   From RSU →

   Accusation = $Ac_{Kp}$ {RR, Sig, TS, AV} (2)

   Sig: is the signature of RSU, $_{Kp}$: public key of CA.

4) Step 4- CA sends an order to RSU, to erase the certificate of the adversary vehicle, and sends new AC for the adversary vehicle.

   From CA →

   Erase= $Er_{Kp}$ {RR, Sig, TS, AV, AC} (3)

   Sig: is the signature of CA, $_{Kp}$: public key of RSU.

5) Step 5- RSU makes a erase order to kill all the keys of the adversary vehicle, and insert AC for it, containing all information required for it, after inserting this certificate in the adversary vehicle, it can't communicate without this certificate, this makes it easy to identify the adversary, after that RSU broadcasts a warning message to all vehicles on the road to add the id of the new adversary at the top of their AL.

   From RSU →

   Erase = $Er_{Kp}$ {RR, Sig, TS, AV} (4)

   From RSU →

   Insert=$ins_{Kp}$ {AC, TS, Sig} (5)

   Sig: is the signature of RSU, $_{Kp}$: public key of AV.

   From RSU →

   Add = $add_{Kp}$ {AV, TS, Sig, RR, RD} (6)

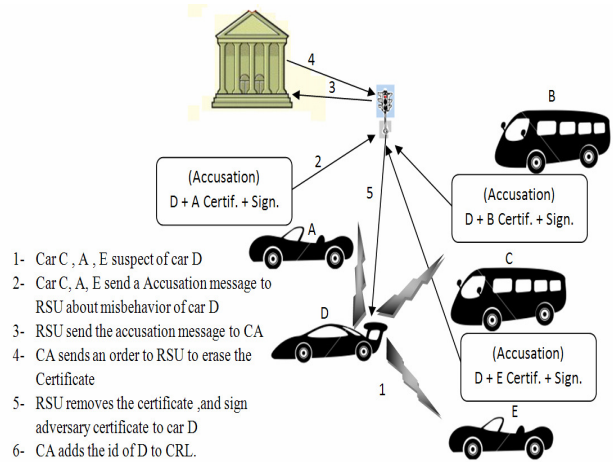

1- Car C, A, E suspect of car D
2- Car C, A, E send a Accusation message to RSU about misbehavior of car D
3- RSU send the accusation message to CA
4- CA sends an order to RSU to erase the Certificate
5- RSU removes the certificate ,and sign adversary certificate to car D
6- CA adds the id of D to CRL.

Fig. 7: VBP Steps.

RD: AC Review Date, $_{Kp}$: Public key for any received vehicle, as this message will be sent to all vehicles on the road.

6) CA adds the id of the new adversary into the CRL.

   Add = add {AV, TS, RR} (7)

When RSU intended to make the revocation, some rules must be considered:

1) Each accuser must attach a "Valid Certificate" with its accusation message; any accusations with "Adversary Certificate" will be ignored.
2) The accusation must contain the accuser signature and must be encrypted with public key of RSU.
3) RSU will consider the accusations if number of accusers are more than threshold related to the total number of the vehicles on the road.

After making the revocation the adversary vehicle will have just one certificate, this certificate tells other vehicles that this vehicle is a trouble maker so "avoid it".

This protocol (VAP) is an optimization of RTPD protocol for revocation, as in RTPD, after erasing all the vehicle certificates, the adversary vehicle can simply contact with any CA, and make request for initiating new certificate, or steal any expired certificate and request for renew, in VAP after erasing all the adversary certificate, AC will be assigned to the adversary so any communication must be through that certificate.

When any vehicle transmits, it includes VC or AC certificate with identity certificate, the receiver will make sure that VC and identity certificate belongs to one sender, identity certificate expires after small period of time, this will prevent the adversary from stealing VC from other vehicles, as it will be difficult to frequently steal more than one certificates from one vehicle or different vehicles due to mobility and other security considerations.

### III. CONCLUSION AND FUTURE WORK

Bad behavior expected to happen frequently in VANET due to the large number of audience, Certificate management process is a critical issue that must be considered to protect the network from possible attacks.

In this paper we introduced new methods for dealing and managing the certificates, these methods will make the network less overhead, and the communication faster, and provides an easier way to recognize the adversary vehicles, in our future work we would like to make simulation for previous protocol and methods.